\documentclass[aps,pra,floatfix,showpacs,preprint,superscriptaddress]{revtex4-1}
\usepackage{bm}
\usepackage{mathrsfs}
\usepackage{amsmath}
\usepackage{amssymb}
\usepackage{revsymb}
\usepackage{accents}
\begin{document}
\title{On the completeness and orthonormality of the Volkov states and the Volkov propagator in configuration space}
\author{A. \surname{Di Piazza}}
\email{dipiazza@mpi-hd.mpg.de}
\affiliation{Max-Planck-Institut f\"ur Kernphysik, Saupfercheckweg 1, D-69117 Heidelberg, Germany}
\date{\today}

\begin{abstract}
Volkov states and Volkov propagator are the basic analytical tools to investigate QED processes occurring in the presence of an intense plane-wave electromagnetic field. In the present paper we provide alternative and relatively simple proofs of the completeness and of the orthonormality at a fixed time of the Volkov states. Concerning the completeness, we exploit some known properties of the Green's function of the Dirac operator in a plane wave, whereas the orthonormality of the Volkov states is proved, relying only on a geometric argument based on the Gauss theorem in four dimensions. In relation with the completeness of the Volkov states, we also study some analytical properties of the Green's function of the Dirac operator in a plane wave, which we explicitly prove to coincide with the Volkov propagator in configuration space. In particular, a closed-form expression in terms of modified Bessel functions and Hankel functions is derived by means of the operator technique in a plane wave and different asymptotic forms are determined. Finally, the transformation properties of the Volkov propagator under general gauge transformations and a general gauge-invariant expression of the so-called dressed mass in configuration space are presented.

  
\end{abstract}
 
\maketitle

\section{Introduction}

The exact solution of the Dirac equation in the presence of a background plane-wave electromagnetic field (indicated below also as ``laser field'') was found by V. D. Volkov well before the invention of the laser \cite{Volkov_1935}. The corresponding one-particle electron states have been widely employed in order to describe quantum electrodynamical processes occurring in the presence of a strong laser field, starting from the pioneering papers by H. R. Reiss \cite{Reiss_1962}, by I. I. Gol'dman \cite{Goldman_1964}, by L. S. Brown and T. W. B. Kibble \cite{Brown_1964}, and by A. I. Nikishov and V. I. Ritus \cite{Nikishov_1964}. In the present context, the ``strength'' of a laser field depends on the value of the so-called classical intensity parameter $\xi_0=|e|E_0/m\omega_0$, where $E_0$ and $\omega_0$ are the electric field amplitude and the central angular frequency of the laser field, respectively, and where $e<0$ and $m$ are the electron charge and mass, respectively (units with $\hbar=c=4\pi\epsilon_0=1$ and $\alpha=e^2\approx 1/137$ are employed throughout). If $\xi_0\gtrsim 1$, in fact, the laser field is able to transfer to an electron an energy corresponding to multiple laser photons in the typical QED length $\lambda_C=1/m=3.9\times 10^{-11}\;\text{cm}$ (Compton wavelength), implying that nonlinear effects in the laser intensity cannot be neglected in the study of the corresponding quantum processes. The inclusion of such nonlinear effects can be carried out from the beginning by treating the laser field as a classical background field and by employing the Volkov states to quantize the electron-positron field. This represents a special application of the so-called Furry picture \cite{Furry_1951,Landau_b_4_1982,Fradkin_b_1991}, which has been widely employed recently, especially due to the development of high-power laser systems. In fact, unprecedented intensities of the order of up to $10^{22}\;\text{W/cm$^2$}$ have already been reported in the literature \cite{Yanovsky_2008} and ultra-strong laser facilities under construction like Apollon 10 PW \cite{APOLLON_10P}, the Extreme Light Infrastructure (ELI) \cite{ELI}, and the Exawatt Center for Extreme Light Studies (XCELS) \cite{XCELS} aim at exceeding the present record by one/two orders of magnitudes. It is worth pointing out, in fact, that for optical laser facilities as those mentioned above, the parameter $\xi_0$ exceeds unity already at laser intensities of the order of $10^{18}\;\text{W/cm$^2$}$. 
Apart from the pioneering papers cited above and by also referring the reader to the reviews \cite{Mitter_1975,Ritus_1985,Ehlotzky_2009,Di_Piazza_2012,Roshchupkin_2012}, we mention here several investigations on nonlinear Compton scattering \cite{Narozhny_2000,Boca_2009,Harvey_2009,Heinzl_2009,Mackenroth_2010,Mackenroth_2011,Seipt_2011,Krajewska_2012,Seipt_2013,Dinu_2013,Nedoreshta_2013,Krajewska_2014,Wistisen_2014,Angioi_2016}, on nonlinear Breit-Wheeler pair production \cite{Narozhny_2000,Roshchupkin_2001,Heinzl_2010b,Mueller_2011b,Titov_2012,Nousch_2012,Krajewska_2013b,Jansen_2013,Augustin_2014}, on nonlinear Bethe-Heitler pair production \cite{Yakovlev_1966,MVG2003PRL,MVG2003PRA,Avetissian_2003,Kaminski_2006,Krajewska_2006,Muller2009PLB}, on electron-positron annihilation \cite{Ilderton_2011b}, and on higher-order processes like nonlinear double Compton scattering \cite{Loetstedt_2009,Seipt_2012,Mackenroth_2013,King_2015} and trident pair production \cite{Hu_2010,Ilderton_2011}. In order to calculate a second-order process like nonlinear double Compton scattering, the corresponding Feynman propagator in a plane-wave field (Volkov propagator) has to be employed \cite{Brown_1964,Reiss_1966,Eberly_1966,Mitter_1975,Ritus_1985,Lavelle_2014,Lavelle_2015}. The latter is also essential to study radiative corrections in the presence of a laser field based on the determination of, e.g., the mass operator \cite{Baier_1976_a} and the polarization operator \cite{Becker_1975,Baier_1976_b} (see also \cite{Meuren_2013}). The representation of the mass operator and of the polarization operator found in \cite{Baier_1976_a} and in \cite{Baier_1976_b}, respectively, exploits the operator technique \cite{Schwinger_1951}, which was later employed to investigate photon splitting in a plane wave \cite{Di_Piazza_2007}. The mass operator and the polarization operator have been later widely used to investigate radiative corrections and vacuum-polarization effects in plane-wave fields \cite{Di_Piazza_2008,Meuren_2011,Di_Piazza_2013,Dinu_2014,Dinu_2014b,Gies_2014} and the total rate of nonlinear Breit-Wheeler pair production via the optical theorem \cite{Milstein_2006} (see also \cite{Meuren_2015}). As an alternative approach to study QED processes in an intense plane wave, still based on the Volkov states, we mention here the Wigner formalism investigated in \cite{Hebenstreit_2011}. Before passing to the results of the present paper, we also recall recent extensions with respect to the Volkov-based approach to study QED processes occurring in tightly focused laser beams \cite{Di_Piazza_2014,Di_Piazza_2015,Di_Piazza_2016,Heinzl_2016,Di_Piazza_2017b,Heinzl_2017,Heinzl_2017b}, in counterpropagating plane waves \cite{King_2016}, and to take into account the back-reaction onto the laser field in QED processes \cite{Ilderton_2018}.

Below, we first study the structure of the Green's function $G(x,x')$ of the Dirac equation in an arbitrary plane wave background field in configuration space. Boundary conditions corresponding to the Feynman prescription $m\to m-i0$ are understood \cite{Landau_b_4_1982}. The Green's function is derived by means of the operator technique as in \cite{Baier_1976_a,Baier_1976_b} and then explicitly expressed in terms of modified Bessel functions and Hankel functions. A direct evaluation of the Green's function in configuration space in terms of the free Green's function $G_0(x,x')$ and its derivatives can be found in \cite{Brown_1964} (see also \cite{Kibble_1975}). Some of the properties of the Green's function $G(x,x')$ like its transformation properties under general gauge transformations and its asymptotic expressions depending on the space-time separation $(x-x')^2$ are also investigated below, where the meaning of the wording ``general gauge transformations'' will be clarified. In addition, an expression of the so-called dressed mass in configuration space is obtained, which is manifestly invariant under a general gauge transformation. Then, the Green's function so obtained is related to the exact Feynman propagator in the plane-wave field (Volkov propagator). By expressing the former function as an integral over the Volkov states with positive- and negative-energy, we show in a relatively straightforward way the completeness of the Volkov states themselves at a fixed time. We also point out that this completeness relation has been already proved by a direct calculation in \cite{Boca_2010}, whereas the completeness of the Volkov states at a fixed light-cone time has been demonstrated in \cite{Bergou_1980}. Finally, we will also prove the orthonormality of the Volkov states at a fixed time by means of a geometrical reasoning, which only exploits the Gauss theorem in four dimensions. It is worth stressing that the orthonormality of Volkov states has also been proved already by a direct calculation in \cite{Ritus_1985,Boca_2010} and, in a mathematically more rigorous way, in \cite{Zakowicz_2005}. We also refer to the recent notes in \cite{Yakaboylu_2015}, where useful details and observations on the derivations in \cite{Ritus_1985,Boca_2010} are presented.

\section{The Green's function of the Dirac equation in a plane wave}

As we have mentioned in the Introduction, the present study focuses on the dynamics of electrons (and positrons) in a plane-wave field. The latter is described by the four-vector potential $A^{\mu}(\phi)$, which only depends on the light-cone time $\phi=t-\bm{n}\cdot \bm{x}$. Here, the unit vector $\bm{n}$ defines the propagation direction of the plane wave, which can be used to introduce two useful four-dimensional quantities: $n^{\mu}=(1,\bm{n})$ and $\tilde{n}^{\mu}=(1,-\bm{n})/2$ (we adopt the metric tensor $\eta^{\mu\nu}=\text{diag}(+1,-1,-1,-1)$ such that $\phi=(nx)$). Assuming obvious differential properties of the four-vector potential $A^{\mu}(\phi)$ and its derivatives, it is clear that it is a solution of the free Maxwell's equation $\square A^{\mu}=0$, where $\square=\partial_{\nu}\partial^{\nu}$, and it is assumed to fulfill the Lorenz-gauge condition $\partial_{\mu}A^{\mu}=0$, with the additional constraint $A^0(\phi)=0$. Thus, if we represent $A^{\mu}(\phi)$ in the form $A^{\mu}(\phi)=(0,\bm{A}(\phi))$, then the Lorenz-gauge condition implies $\bm{n}\cdot\bm{A}'(\phi)=0$, with the prime indicating the derivative with respect to $\phi$. If we make the physically reasonable assumption that $\bm{A}(\phi)$ vanishes for $\phi\to\pm\infty$, the equality $\bm{n}\cdot\bm{A}'(\phi)=0$ implies that $\bm{n}\cdot\bm{A}(\phi)=0$. By introducing two four-vectors $a_j^{\mu}=(0,\bm{a}_j)$, with $j=1,2$, such that $(na_j)=-\bm{n}\cdot\bm{a}_j=0$ and $(a_ia_j)=-\bm{a}_i\cdot\bm{a}_j=-\delta_{ij}$, the most general form of the vector potential $\bm{A}(\phi)$ reads $\bm{A}(\phi)=\psi_1(\phi)\bm{a}_1+\psi_2(\phi)\bm{a}_2$, where the two functions $\psi_j(\phi)$ are arbitrary provided that they vanish for $\phi\to\pm\infty$ and they feature the differential properties mentioned above when the four-vector potential $A^{\mu}(\phi)$ was introduced. The four-dimensional quantities $n^{\mu}$, $\tilde{n}^{\mu}$, and $a^{\mu}_j$ fulfill the completeness relation: $\eta^{\mu\nu}=n^{\mu}\tilde{n}^{\nu}+\tilde{n}^{\mu}n^{\nu}-a_1^{\mu}a_1^{\nu}-a_2^{\mu}a_2^{\nu}$ (note that $(n\tilde{n})=1$ and $(\tilde{n}a_j)=0$). Below, we will refer to the longitudinal ($n$) direction as the direction along $\bm{n}$ and to the transverse ($\perp$) plane as the plane spanned by the two perpendicular unit vectors $\bm{a}_j$. In this respect, together with the light-cone time $\phi=t-x_n$, with $x_n=\bm{n}\cdot \bm{x}$, we also introduce the remaining three light-cone coordinates $T=(\tilde{n}x)=(t+x_n)/2$, and $\bm{x}_{\perp}=(x_{a_1},x_{a_2})=(\bm{x}\cdot\bm{a}_1,\bm{x}\cdot\bm{a}_2)$. Analogously, the light-cone coordinates of an arbitrary four-vector $v^{\mu}=(v^0,\bm{v})$ will be indicated as $v_-=(nv)=v^0-v_n$, with $v_n=\bm{n}\cdot \bm{v}$, $v_+=(\tilde{n}v)=(v^0+v_n)/2$, and $\bm{v}_{\perp}=(v_{a_1},v_{a_2})=(\bm{v}\cdot\bm{a}_1,\bm{v}\cdot\bm{a}_2)$. Since we will employ the operator technique, it is convenient to also introduce the momenta operators $P_{\phi}=-i\partial_{\phi}=-(\tilde{n}P)=-(i\partial_t-i\partial_{x_n})/2$, $P_T=-i\partial_T=-(nP)=-(i\partial_t+i\partial_{x_n})$, and $\bm{P}_{\perp}=(P_{a_1},P_{a_2})=-i(\bm{a}_1\cdot\bm{\nabla},\bm{a}_2\cdot\bm{\nabla})$. These operators are the momenta conjugated to the light-cone coordinates in the sense that the commutator between the operator corresponding to each light-cone coordinate and the associated momentum operator is equal to the imaginary unit (all other possible commutators vanish). Note that if $|x\rangle$ ($|p\rangle$) is the eigenstate of the four-position (four-momentum) operator $X^{\mu}$ ($P^{\mu}=i\partial^{\mu}$) with eigenvalue $x^{\mu}$ ($p^{\mu}$), i.e., $X^{\mu}|x\rangle=x^{\mu}|x\rangle$ ($P^{\mu}|p\rangle=p^{\mu}|p\rangle$), then, by normalizing the eigenstates $|x\rangle$ ($|p\rangle$) such that $\langle x|y\rangle=\delta^{(4)}(x-y)$ [$\langle p|q\rangle=(2\pi)^4\delta^{(4)}(p-q)$], it is $\langle x|p\rangle=\exp(-i(px))=\exp[-i(p_+\phi+p_-T-\bm{p}_{\perp}\cdot\bm{x}_{\perp})]$ and $P_{\phi}|p\rangle=-p_+|p\rangle$, $P_T|p\rangle=-p_-|p\rangle$, and $\bm{P}_{\perp}|p\rangle=\bm{p}_{\perp}|p\rangle$.

The electron Green's function $G_b(x,x')$ in a general background electromagnetic field described by the four-vector potential $A_b^{\mu}(x)$ is defined by the equation
\begin{equation}
\label{Eq_G_b}
\{\gamma^{\mu}[i\partial_{\mu}-eA_{b,\mu}(x)]-m\}G_b(x,x')=\delta^{(4)}(x-x'),
\end{equation}
where $\gamma^{\mu}$ are the Dirac matrices satisfying the anti-commutation relations $\{\gamma^{\mu},\gamma^{\nu}\}=2\eta^{\mu\nu}$ \cite{Landau_b_4_1982}. In order to uniquely identify the Green's function, boundary conditions have also to be specified. When passing to the four-momentum ($q^{\mu}$) representation for the computation of the Green's function, this corresponds to the necessity of shifting the poles at $q^0=\pm\sqrt{m^2+\bm{q}^2}$ of the four-dimensional Fourier transform of the Green's function (depending on the structure of the external field other poles may be present). Here, we always assume the Feynman prescription corresponding to the shift $m\to m-i0$ \cite{Landau_b_4_1982}. Moreover, under the gauge transformation $A^{\mu}(x)\to A^{\mu}(x)+\partial^{\mu}\Lambda(x)$, the Green's function $G_b(x,x')$ transforms according to the rule \cite{Landau_b_4_1982}
\begin{equation}
\label{GT}
G_b(x,x')\to e^{-ie[\Lambda(x)-\Lambda(x')]}G_b(x,x').
\end{equation}
Within the operator technique the operator $G_b$ corresponding to the Green's function $G_b(x,x')$ is defined via the equation $G_b(x,x')=\langle x|G_b|x'\rangle$, i.e., as [see Eq. (\ref{Eq_G_b})]
\begin{equation}
G_b=\frac{1}{\hat{\Pi}_b-m+i0},
\end{equation}
where $\hat{v}=\gamma^{\mu}v_{\mu}$ for a generic four-vector $v^{\mu}$ and where $\Pi_b^{\mu}=P^{\mu}-eA_b^{\mu}(X)$. The transformation of the Green's function under a gauge transformation is easily obtained within the operator technique. In fact, from the commutation relation between the four-position operators and the four-momentum operators, it is clear that $[P^{\mu},f(X)]=i[\partial^{\mu}f(X)]$, where $f(X)$ is an arbitrary function of the four-position operator (with an abuse of notation, we have indicated as $\partial^{\mu}$ the partial derivative with respect to the operator $X_{\mu}$). Analogously, it can easily be shown that $\exp[if(X)]P^{\mu}\exp[-if(X)]=P^{\mu}+\partial^{\mu}f(X)$ and, more in general, $\exp[if(X)]g(P)\exp[-if(X)]=g(P+\partial f(X))$, for an arbitrary function $g(P)$ of the four-momentum operators. In this way, if $G'_b$ is the Green's function corresponding to the gauge-transformed field $A^{\mu}_b(X)+\partial^{\mu}\Lambda(X)$, i.e., if
\begin{equation}
G'_b=\frac{1}{\hat{\Pi}'_b-m+i0},
\end{equation}
with $\Pi^{\prime\,\mu}_b=\Pi_b^{\mu}-e\partial^{\mu}\Lambda(X)$, it is
\begin{equation}
G'_b=e^{-ie\Lambda(X)}\frac{1}{\hat{\Pi}_b-m+i0}e^{ie\Lambda(X)},
\end{equation}
which provides the correct transformation law once the matrix element $\langle x|G'_b|x'\rangle$ is evaluated.

From now on we focus on the background field being the already introduced plane wave: $A^{\mu}_b(x)=A^{\mu}(\phi)$, with the corresponding operator $\Pi^{\mu}=P^{\mu}-eA^{\mu}(\Phi)$ and Green's function operator $G=(\hat{\Pi}-m+i0)^{-1}$ [$G(x,x')=\langle x|G|x'\rangle$]. By employing the usual Schwinger representation \cite{Schwinger_1951}, we have
\begin{equation}
G=\frac{1}{\hat{\Pi}-m+i0}=(\hat{\Pi}+m)\frac{1}{\hat{\Pi}^2-m^2+i0}=(-i)(\hat{\Pi}+m)\int_0^{\infty}ds\, e^{is(\hat{\Pi}^2-m^2)},
\end{equation}
where, starting from the last equality on, the prescription $m^2\to m^2-i0$ is understood. Now, it is easily shown from the anti-commutation relations of the $\gamma$-matrices that
\begin{equation}
\hat{\Pi}^2-m^2=[P-eA(\Phi)]^2-m^2-\frac{ie}{2}\sigma^{\mu\nu}F_{\mu\nu}(\Phi)=2P_TP_{\phi}-[\bm{P}_{\perp}-e\bm{A}_{\perp}(\Phi)]^2-m^2-ie\hat{n}\hat{A}'(\Phi),
\end{equation}
where $\sigma^{\mu\nu}=[\gamma^{\mu},\gamma^{\nu}]/2$, $F_{\mu\nu}(\Phi)=\partial_{\mu}A_{\nu}(\Phi)-\partial_{\nu}A_{\mu}(\Phi)=n_{\mu}A'_{\nu}(\Phi)-n_{\nu}A'_{\mu}(\Phi)$ (here, again with an abuse of notation, the prime indicates the derivative with respect to the operator $\Phi$ corresponding to the coordinate $\phi$). In an analogous way as we have indicated, e.g., in \cite{Di_Piazza_2015}, we have to disentangle now the operator $\exp(is\{2P_TP_{\phi}-[\bm{P}_{\perp}-e\bm{A}_{\perp}(\Phi)]^2-m^2-ie\hat{n}\hat{A}'(\Phi)\})$, by writing it in the form
\begin{equation}
\label{Disentanglement}
e^{is\{2P_TP_{\phi}-[\bm{P}_{\perp}-e\bm{A}_{\perp}(\Phi)]^2-m^2-ie\hat{n}\hat{A}'(\Phi)\}}=L(s)e^{2isP_TP_{\phi}}.
\end{equation}
In order to determine the operator $L(s)$, we observe that it satisfies the differential equation
\begin{equation}
\label{dOds}
\begin{split}
\frac{dL}{ds}=&-iLe^{2isP_TP_{\phi}}\{[\bm{P}_{\perp}-e\bm{A}_{\perp}(\Phi)]^2+m^2+ie\hat{n}\hat{A}'(\Phi)\}e^{-2isP_TP_{\phi}}\\
=&-iL\{[\bm{P}_{\perp}-e\bm{A}_{\perp}(\Phi+2sP_T)]^2+m^2+ie\hat{n}\hat{A}'(\Phi+2sP_T)\},
\end{split}
\end{equation}
where we have used the fact that, since $[\Phi,P_{\phi}]=i$ and $[\Phi,P_T]=0$, then for an arbitrary function $f(\Phi)$ it is $\exp(2isP_{\phi}P_T)f(\Phi)\exp(-2isP_{\phi}P_T)=f(\Phi+2sP_T)$. The solution of Eq. (\ref{dOds}), with the initial condition $L(0)=1$ [see Eq. (\ref{Disentanglement})], is
\begin{equation}
L(s)=e^{-i\int_0^sds'\{[\bm{P}_{\perp}-e\bm{A}_{\perp}(\Phi+2s'P_T)]^2+m^2+ie\hat{n}\hat{A}'(\Phi+2s'P_T)\}}.
\end{equation}
Since all operators in the exponent of $L(s)$ now commute with each other and since in general $\hat{n}\hat{A}(\Phi)\hat{n}\hat{A}(\Phi')=0$, we have
\begin{equation}
\label{G}
\begin{split}
G=&(-i)(\hat{\Pi}+m)\int_0^{\infty}ds\, e^{-im^2s}\Big\{1+\frac{e}{2P_T}\hat{n}[\hat{A}(\Phi+2sP_T)-\hat{A}(\Phi)]\Big\}\\
&\times e^{-i\int_0^sds'[\bm{P}_{\perp}-e\bm{A}_{\perp}(\Phi+2s'P_T)]^2}e^{2isP_TP_{\phi}},
\end{split}
\end{equation}
which coincides with the corresponding expression in \cite{Baier_1976_a,Baier_1976_b} (see also \cite{Di_Piazza_2007}). Now, by writing $|x\rangle=|\phi,T,\bm{x}_{\perp}\rangle$ and $|x'\rangle=|\phi',T',\bm{x}'_{\perp}\rangle$, we obtain
\begin{equation}
\begin{split}
G(x,x')=&(-i)\left[i\hat{n}\partial_{\phi}+i\hat{\tilde{n}}\partial_T+i\bm{\gamma}_{\perp}\cdot\bm{\nabla}_{\perp}-e\hat{A}(\phi)+m\right]\int_0^{\infty}ds\, e^{-im^2s}\\
&\times\langle\phi,T,\bm{x}_{\perp}|\Big\{1+\frac{e}{2P_T}\hat{n}[\hat{A}(\Phi+2sP_T)-\hat{A}(\Phi)]\Big\}\\
&\times e^{-i\int_0^sds'[\bm{P}_{\perp}-e\bm{A}_{\perp}(\Phi+2s'P_T)]^2}e^{2isP_TP_{\phi}}|\phi',T',\bm{x}'_{\perp}\rangle.
\end{split}
\end{equation}
Since the operator between the two four-position eigenstates does not contain $P_{\phi}$, it is relatively easy to evaluate the corresponding matrix element by noticing that $\exp(2isP_TP_{\phi})|\phi\rangle=|\phi-2sP_T\rangle$. The result is
\begin{equation}
\begin{split}
G(x,x')=&(-i)\left[i\hat{n}\partial_{\phi}+i\hat{\tilde{n}}\partial_T+i\bm{\gamma}_{\perp}\cdot\bm{\nabla}_{\perp}-e\hat{A}(\phi)+m\right]\int_0^{\infty}ds\, e^{-im^2s}\\
&\times\langle T,\bm{x}_{\perp}|\Big\{1+\frac{e}{2P_T}\hat{n}[\hat{A}(\phi+2sP_T)-\hat{A}(\phi)]\Big\}\\
&\times e^{-i\int_0^sds'[\bm{P}_{\perp}-e\bm{A}_{\perp}(\phi+2s'P_T)]^2}\delta(\phi-\phi'+2sP_T)|T',\bm{x}'_{\perp}\rangle.
\end{split}
\end{equation}
By inserting the unity operator $\int dp_- d^2p_{\perp}(2\pi)^{-3}|p_-,\bm{p}_{\perp}\rangle\langle p_-,\bm{p}_{\perp}|$ and by exploiting the $\delta$-function to take the integral in $p_-$, we have
\begin{equation}
\begin{split}
G(x,x')=&-\frac{i}{4\pi}\left[i\hat{n}\partial_{\phi}+i\hat{\tilde{n}}\partial_T+i\bm{\gamma}_{\perp}\cdot\bm{\nabla}_{\perp}-e\hat{A}(\phi)+m\right]\int_0^{\infty}\frac{ds}{s}\, e^{-im^2s}\\
&\times\int\frac{d^2p_{\perp}}{(2\pi)^2}\bigg[1+es\hat{n}\frac{\hat{A}(\phi')-\hat{A}(\phi)}{\phi'-\phi}\bigg]e^{-i(T-T')(\phi-\phi')/2s+i\bm{p}_{\perp}\cdot(\bm{x}_{\perp}-\bm{x}'_{\perp})}\\
&\times e^{-i\int_0^sds'[\bm{p}_{\perp}-e\bm{A}_{\perp}(\phi+s'(\phi'-\phi)/s)]^2}.
\end{split}
\end{equation}
The integral in $\bm{p}_{\perp}$ is Gaussian and can be taken by employing the general formula \cite{Gradshteyn_b_2000}
\begin{equation}
\int_{-\infty}^{\infty} dx\, e^{-iax^2}=e^{-i\pi/4}\sqrt{\frac{\pi}{a}}
\end{equation}
for $a>0$, and the final result is
\begin{equation}
\begin{split}
G(x,x')=&-\frac{i}{16\pi^2}\left[i\hat{n}\partial_{\phi}+i\hat{\tilde{n}}\partial_T+i\bm{\gamma}_{\perp}\cdot\bm{\nabla}_{\perp}-e\hat{A}(\phi)+m\right]e^{-ie(x-x')^{\mu}\int_0^1d\lambda A_{\mu}(\phi'+\lambda(\phi-\phi'))}\\
&\times\int_0^{\infty}\frac{ds}{s^2}\,\bigg[1+es\hat{n}\frac{\hat{A}(\phi)-\hat{A}(\phi')}{\phi-\phi'}\bigg]e^{-is\tilde{m}^2(x,x')-i(x-x')^2/4s},
\end{split}
\end{equation}
where
\begin{equation}
\label{m_tilde}
\tilde{m}^2(x,x')=m^2-e^2\int_0^1d\lambda A^2(\phi'+\lambda(\phi-\phi'))+e^2\left[\int_0^1d\lambda A^{\mu}(\phi'+\lambda(\phi-\phi'))\right]^2,
\end{equation}
with $\tilde{m}^2(x,x')\ge m^2$, is the square of the well-known dressed mass in configuration space \cite{Brown_1964,Kibble_1975,Hebenstreit_2011,Di_Piazza_2015}. The reason why we have superfluously indicated the dressed mass as dependent on all coordinates, will be clear below.

The remaining integral on the proper time $s$ depends on the sign of $(x-x')^2$. First, we notice that
\begin{equation}
\label{I_2}
I_2=\int_0^{\infty}\frac{ds}{s^2}\,e^{-is\tilde{m}^2(x,x')-i(x-x')^2/4s}=4i\frac{\partial I_1}{\partial[(x-x')^2]},
\end{equation}
where
\begin{equation}
I_1=\int_0^{\infty}\frac{ds}{s}\,e^{-is\tilde{m}^2(x,x')-i(x-x')^2/4s}.
\end{equation}
Now, if $(x-x')^2>0$, we can set $\tilde{m}^2(x,x')s=B_+(x,x')\rho$ and $(x-x')^2/4s=B_+(x,x')/\rho$, which implies $B_+(x,x')=\tilde{m}(x,x')\sqrt{(x-x')^2}/2$, whereas if $(x-x')^2<0$, we can set $\tilde{m}^2(x,x')s=B_-(x,x')\rho$ and $(x-x')^2/4s=-B_-(x,x')/\rho$, which implies $B_-(x,x')=\tilde{m}(x,x')\sqrt{-(x-x')^2}/2$. In this way, we have
\begin{equation}
I_1=\theta\big((x-x')^2\big)\int_0^{\infty}\frac{d\rho}{\rho}\,e^{-iB_+(x,x')(\rho+1/\rho)}+\theta\big(-(x-x')^2\big)\int_0^{\infty}\frac{d\rho}{\rho}\,e^{-iB_-(x,x')(\rho-1/\rho)}.
\end{equation}
These integrals can be expressed in terms of the Hankel function $\text{H}^{(2)}_0(z)$ and of the modified Bessel function $\text{K}_0(z)$ \cite{Gradshteyn_b_2000} as
\begin{equation}
I_1=-i\pi\theta\big((x-x')^2\big)\text{H}^{(2)}_0\big(\tilde{m}(x,x')\sqrt{(x-x')^2}\big)+2\theta\big(-(x-x')^2\big)\text{K}_0\big(\tilde{m}(x,x')\sqrt{-(x-x')^2}\big).
\end{equation}
Now, in the limit of vanishing argument, the asymptotic expansions hold \cite{Gradshteyn_b_2000}
\begin{align}
\text{H}^{(2)}_0(z)&\sim 1-i\frac{2}{\pi}\left[\log\left(\frac{z}{2}\right)+C\right] && z\to 0,\\
\text{K}_0(z)&\sim -\left[\log\left(\frac{z}{2}\right)+C\right] && z\to 0,
\end{align}
where $C=0.577...$ is the Euler constant. Thus, from Eq. (\ref{I_2}) we have
\begin{equation}
\begin{split}
I_2=&4\pi\delta\big((x-x')^2\big)-2\pi\theta\big((x-x')^2\big)\frac{\tilde{m}(x,x')}{\sqrt{(x-x')^2}}\text{H}^{(2)}_1\big(\tilde{m}(x,x')\sqrt{(x-x')^2}\big)\\
&+4i\theta\big(-(x-x')^2\big)\frac{\tilde{m}(x,x')}{\sqrt{-(x-x')^2}}\text{K}_1\big(\tilde{m}(x,x')\sqrt{-(x-x')^2}\big).
\end{split}
\end{equation}
In conclusion, we obtain the explicit expression of the Green's function in the form
\begin{equation}
\begin{split}
G(x,x')=&-\frac{1}{4\pi}\left[i\hat{n}\partial_{\phi}+i\hat{\tilde{n}}\partial_T+i\bm{\gamma}_{\perp}\cdot\bm{\nabla}_{\perp}-e\hat{A}(\phi)+m\right]e^{-ie(x-x')^{\mu}\int_0^1d\lambda A_{\mu}(\phi'+\lambda(\phi-\phi'))}\\
&\times\left\{\delta\big((x-x')^2\big)-\frac{1}{2}\theta\big((x-x')^2\big)\frac{\tilde{m}(x,x')}{\sqrt{(x-x')^2}}\text{H}^{(2)}_1\big(\tilde{m}(x,x')\sqrt{(x-x')^2}\big)\right.\\
&+\frac{i}{\pi}\theta\big(-(x-x')^2\big)\frac{\tilde{m}(x,x')}{\sqrt{-(x-x')^2}}\text{K}_1\big(\tilde{m}(x,x')\sqrt{-(x-x')^2}\big)\\
&+\frac{e}{4\pi}\hat{n}\frac{\hat{A}(\phi)-\hat{A}(\phi')}{\phi-\phi'}\left[-i\pi\theta\big((x-x')^2\big)\text{H}^{(2)}_0\big(\tilde{m}(x,x')\sqrt{(x-x')^2}\big)\right.\\
&\left.+2\theta\big(-(x-x')^2\big)\text{K}_0\big(\tilde{m}(x,x')\sqrt{-(x-x')^2}\big)\right]\Bigg\}.
\end{split}
\end{equation}
This expression of the Green's function can be significantly simplified by observing that (see, e.g., \cite{Bogoliubov_b_1980})
\begin{align}
\begin{split}
\sqrt{(x-x')^2-i0}=&\theta\big((x-x')^2\big)\sqrt{(x-x')^2}\\
&-i\theta\big(-(x-x')^2\big)\sqrt{-(x-x')^2},
\end{split}\\
\begin{split}
\text{H}^{(2)}_0\big(\tilde{m}(x,x')\sqrt{(x-x')^2-i0}\big)=&\theta\big((x-x')^2\big)\text{H}^{(2)}_0\big(\tilde{m}(x,x')\sqrt{(x-x')^2}\big)\\
&+i\frac{2}{\pi}\theta\big(-(x-x')^2\big)\text{K}_0\big(\tilde{m}(x,x')\sqrt{-(x-x')^2}\big),
\end{split}\\
\begin{split}
\frac{\text{H}^{(2)}_1\big(\tilde{m}(x,x')\sqrt{(x-x')^2-i0}\big)}{\sqrt{(x-x')^2-i0}}=&\theta\big((x-x')^2\big)\frac{\text{H}^{(2)}_1\big(\tilde{m}(x,x')\sqrt{(x-x')^2}\big)}{\sqrt{(x-x')^2}}\\
&-i\frac{2}{\pi}\theta\big(-(x-x')^2\big)\frac{\text{K}_1\big(\tilde{m}(x,x')\sqrt{-(x-x')^2}\big)}{\sqrt{-(x-x')^2}}.
\end{split}
\end{align}
By exploiting these identities, in fact, we can write the Green's function simply as (see also \cite{Brown_1964,Bogoliubov_b_1980,Di_Piazza_2015})
\begin{equation}
\label{G_f}
\begin{split}
G(x,x')=&-\frac{1}{4\pi}e^{-ie(x-x')^{\mu}\int_0^1d\lambda A_{\mu}(\phi'+\lambda(\phi-\phi'))}\\
&\times\gamma^{\rho}\left[i\partial_{\rho}+e\int_0^1d\lambda\,\lambda F_{\rho\sigma}(\phi'+\lambda(\phi-\phi'))(x-x')^{\sigma}+m\right]\\
&\times\bigg[\delta\big((x-x')^2\big)-\frac{1}{2}\frac{\tilde{m}(x,x')}{\sqrt{(x-x')^2-i0}}\text{H}^{(2)}_1\big(\tilde{m}(x,x')\sqrt{(x-x')^2-i0}\big)\\
&\quad\left.-i\frac{e}{8}\sigma^{\alpha\beta}\int_0^1d\lambda\,F_{\alpha\beta}(\phi'+\lambda(\phi-\phi'))\text{H}^{(2)}_0\big(\tilde{m}(x,x')\sqrt{(x-x')^2-i0}\big)\right].
\end{split}
\end{equation}
where we have used the operator identity \cite{Brown_1964}
\begin{equation}
\begin{split}
&\left[i\hat{n}\partial_{\phi}+i\hat{\tilde{n}}\partial_T+i\bm{\gamma}_{\perp}\cdot\bm{\nabla}_{\perp}-e\hat{A}(\phi)\right]e^{-ie(x-x')^{\mu}\int_0^1d\lambda A_{\mu}(\phi'+\lambda(\phi-\phi'))}\\
&\quad=\gamma^{\rho}[i\partial_{\rho}-eA_{\rho}(\phi)]e^{-ie(x-x')^{\mu}\int_0^1d\lambda A_{\mu}(\phi'+\lambda(\phi-\phi'))}\\
&\quad=e^{-ie(x-x')^{\mu}\int_0^1d\lambda A_{\mu}(\phi'+\lambda(\phi-\phi'))}\gamma^{\rho}\left[i\partial_{\rho}+e\int_0^1d\lambda\,\lambda F_{\rho\sigma}(\phi'+\lambda(\phi-\phi'))(x-x')^{\sigma}\right].
\end{split}
\end{equation}

\subsection{Gauge transformation of the Green's function and the dressed mass}

The expression in Eq. (\ref{G_f}) is in agreement with the one in \cite{Brown_1964} and is particularly useful to show that the Green's function $G(x,x')$ transforms correctly under a general gauge transformations: $A^{\mu}(\phi)\to A^{\mu}(\phi)+\partial^{\mu}\Lambda(x)$, i.e., a gauge transformation which does not necessarily keeps the four-vector potential to depend only on $\phi$. In fact, it is clear that 
\begin{equation}
\begin{split}
e^{-ie(x-x')^{\mu}\int_0^1d\lambda A_{\mu}(\phi'+\lambda(\phi-\phi'))}&\to e^{-ie(x-x')^{\mu}\int_0^1d\lambda A_{\mu}(\phi'+\lambda(\phi-\phi'))}e^{-ie(x-x')^{\mu}\int_0^1d\lambda \nabla_{\mu}\Lambda(x'+\lambda(x-x'))}\\
&=e^{-ie[\Lambda(x)-\Lambda(x')]}e^{-ie(x-x')^{\mu}\int_0^1d\lambda A_{\mu}(\phi'+\lambda(\phi-\phi'))},
\end{split}
\end{equation}
where the symbol $\nabla_{\mu}$ indicates the partial derivative with respect to the contravariant $\mu$-component of the argument of the function $\Lambda(x)$ and where we have used the relation $(x-x')^{\mu}\nabla_{\mu}\Lambda(x'+\lambda(x-x'))=d\Lambda(x'+\lambda(x-x'))/d\lambda$. In this way, the Green's function $G(x,x')$ will correctly transform as in Eq. (\ref{GT}) if we show that the dressed mass $\tilde{m}(x,x')$ is gauge invariant. This result has been hinted in \cite{Kibble_1975} (see also \cite{Hebenstreit_2011}) but, to the best of the author's knowledge, it has only been explicitly shown for a special gauge transformation, which keeps the four-vector potential to depend only on $\phi$, i.e., with $\Lambda(x)=\tilde{\Lambda}(\phi)$ such that $\partial^{\mu}\Lambda(x)=n^{\mu}\tilde{\Lambda}'(\phi)$. Now, it is clear that [see Eq. (\ref{m_tilde})]
\begin{equation}
\begin{split}
\tilde{m}^2(x,x')=&m^2-e^2\int_0^1d\lambda [A^{\mu}(\phi'+\lambda(\phi-\phi'))-A^{\mu}(\phi')]^2\\
&+e^2\left\{\int_0^1d\lambda [A^{\mu}(\phi'+\lambda(\phi-\phi'))-A^{\mu}(\phi')]\right\}^2
\end{split}
\end{equation}
and that
\begin{equation}
\label{A_F}
\begin{split}
&A^{\mu}(\phi'+\lambda(\phi-\phi'))-A^{\mu}(\phi')=\int_0^{\lambda}d\lambda'\frac{d}{d\lambda'}A^{\mu}(\phi'+\lambda'(\phi-\phi'))\\
&\quad=\int_0^{\lambda}d\lambda'(x-x')_{\nu}\nabla^{\nu}A^{\mu}(\phi'+\lambda'(\phi-\phi'))\\
&\quad=\mathcal{F}^{\mu}(x,x';\lambda)+n^{\mu}(x-x')_{\nu}\int_0^{\lambda}d\lambda'A^{\prime\,\nu}(\phi'+\lambda'(\phi-\phi')),
\end{split}
\end{equation}
where we have introduced the gauge-invariant four-vector
\begin{equation}
\mathcal{F}^{\mu}(x,x';\lambda)=(x-x')_{\nu}\int_0^{\lambda}d\lambda'F^{\nu\mu}(\phi'+\lambda'(\phi-\phi')).
\end{equation}
Since $n^2=(n\mathcal{F}(x,x';\lambda))=0$, we have that
\begin{equation}
\label{m_tilde_F}
\tilde{m}^2(x,x')=m^2-e^2\int_0^1d\lambda\, \mathcal{F}^2(x,x';\lambda)
+e^2\left[\int_0^1d\lambda\, \mathcal{F}^{\mu}(x,x';\lambda)\right]^2,
\end{equation}
which is also manifestly gauge invariant. Before concluding this section, we report another
manifestly gauge-invariant expression of the dressed mass, which also elucidates the physical 
meaning of this quantity. We start from the exact expression of the on-shell kinetic four-momentum $\pi^{\mu}(\phi)$ 
of an electron in a plane wave at a given $\phi$ in terms of the same quantity $\pi^{\mu}(\phi')$ at another
$\phi'$ (see, e.g., \cite{Landau_b_2_1975,Di_Piazza_2012}):
\begin{equation}
\label{pi^mu}
\begin{split}
\pi^{\mu}(\phi)&=\pi^{\mu}(\phi')-e[A^{\mu}(\phi)-A^{\mu}(\phi')]+\frac{e}{\pi_-(\phi')}[A^{\nu}(\phi)-A^{\nu}(\phi')]\pi_{\nu}(\phi')n^{\mu}\\
&\quad-\frac{e^2}{2\pi_-(\phi')}[A^{\nu}(\phi)-A^{\nu}(\phi')]^2n^{\mu}\\
&=\pi^{\mu}(\phi')+\frac{e}{\pi_-(\phi')}\mathscr{F}^{\mu\nu}(\phi,\phi')\pi_{\nu}(\phi')+\frac{e^2}{2\pi_-^2(\phi')}\mathscr{F}^{\mu\nu}(\phi,\phi')\mathscr{F}_{\nu\lambda}(\phi,\phi')\pi^{\lambda}(\phi'),
\end{split}
\end{equation}
where the gauge-invariant quantity $\mathscr{F}^{\mu\nu}(\phi,\phi')=\int_{\phi'}^{\phi}d\varphi F^{\mu\nu}(\varphi)=(\phi-\phi')\int_0^1d\lambda F^{\mu\nu}(\phi'+\lambda(\phi-\phi'))$ has been introduced. Now, it is easy to show that, if we define the average $\langle f\rangle(\phi,\phi')=(\phi-\phi')^{-1}\int_{\phi'}^{\phi}d\varphi f(\varphi)=\int_0^1d\lambda f(\phi'+\lambda(\phi-\phi'))$ of a generic function $f(\phi)$, then
\begin{equation}
\begin{split}
\langle\pi^{\mu}\rangle(\phi,\phi')=&\pi^{\mu}(\phi')+\frac{e}{\pi_-(\phi')}n^{\mu}\int_0^1d\lambda [A^{\nu}(\phi'+\lambda(\phi-\phi'))-A^{\nu}(\phi')]\pi_{\nu}(\phi')\\
&-e\int_0^1d\lambda [A^{\mu}(\phi'+\lambda(\phi-\phi'))-A^{\mu}(\phi')]\\
&-\frac{e^2}{2\pi_-(\phi')}n^{\mu}\int_0^1d\lambda [A^{\nu}(\phi'+\lambda(\phi-\phi'))-A^{\nu}(\phi')]^2
\end{split}
\end{equation}
and we obtain (see also \cite{Hebenstreit_2011,Di_Piazza_2013}) 
\begin{equation}
\label{m_tilde_pi}
\tilde{m}^2(x,x')=\langle\pi^{\mu}\rangle(\phi,\phi')\langle\pi_{\mu}\rangle(\phi,\phi')=m^2-\frac{1}{\phi-\phi'}\int_{\phi'}^{\phi}d\varphi[\pi^{\mu}(\varphi)-\langle\pi^{\mu}\rangle(\phi,\phi')]^2.
\end{equation}
Thus, the square $\tilde{m}^2(x,x')$ of the dressed mass coincides with the square of the average kinetic four-momentum of the electron in the plane wave between $\phi$ and $\phi'$.

\subsection{Asymptotic properties of the Green's function}

The asymptotic expressions of the Green's function $G(x,x')$ for small and large values of the absolute value of $(x-x')^2$ can be found starting from the corresponding asymptotic behavior of the modified Bessel functions and of the Hankel functions. For small space-time intervals, we recall that \cite{Gradshteyn_b_2000}
\begin{align}
\text{H}^{(2)}_0(z)&\sim 1-i\frac{2}{\pi}\left[\log\left(\frac{z}{2}\right)+C\right] && z\to 0,\\
\text{K}_0(z)&\sim -\left[\log\left(\frac{z}{2}\right)+C\right] && z\to 0,\\
\text{H}^{(2)}_1(z)&\sim \frac{z}{2}+i\frac{2}{\pi}\left\{\frac{1}{z}-\frac{z}{2}\left[\log\left(\frac{z}{2}\right)+C-\frac{1}{2}\right]\right\} && z\to 0,\\
\text{K}_1(z)&\sim \frac{1}{z}+\frac{z}{2}\left[\log\left(\frac{z}{2}\right)+C-\frac{1}{2}\right] && z\to 0.
\end{align}
Thus, in the corresponding limit $(x-x')^2\to 0$ we obtain
\begin{equation}
\begin{split}
G(x,x')\sim &\frac{i}{4\pi^2}e^{-ie(x-x')^{\mu}\int_0^1d\lambda A_{\mu}(\phi'+\lambda(\phi-\phi'))}\\
&\times\gamma^{\rho}\left[i\partial_{\rho}+e\int_0^1d\lambda\,\lambda F_{\rho\sigma}(\phi'+\lambda(\phi-\phi'))(x-x')^{\sigma}+m\right]\bigg\{\frac{1}{(x-x')^2-i0}\\
&\qquad-\frac{1}{2}\tilde{m}^2(x,x')\left[\log\left(\frac{\tilde{m}(x,x')\sqrt{(x-x')^2-i0}}{2}\right)+C-\frac{1}{2}+i\frac{\pi}{2}\right]\\
&\qquad-i\frac{e}{4}\sigma^{\alpha\beta}\int_0^1d\lambda\,F_{\alpha\beta}(\phi'+\lambda(\phi-\phi'))\\
&\qquad\qquad\left.\times\left[\log\left(\frac{\tilde{m}(x,x')\sqrt{(x-x')^2-i0}}{2}\right)+C+i\frac{\pi}{2}\right]\right\}.
\end{split}
\end{equation}
Concerning the case of large space-time intervals $|(x-x')^2|$, we distinguish the two sub-cases of large, time-like intervals: $(x-x')^2\to\infty$ and of large, space-like intervals $(x-x')^2\to-\infty$. In the former sub-case we have that \cite{Gradshteyn_b_2000}
\begin{align}
\text{H}^{(2)}_{\nu}(z)\sim\sqrt{\frac{2}{\pi z}}e^{-i[z-(2\nu+1)/4]} && z\to \infty,
\end{align}
and then, in the corresponding limit $(x-x')^2\to\infty$, we obtain
\begin{equation}
\begin{split}
G(x,x')\sim &\frac{i}{8\pi}e^{i\pi/4}\sqrt{\frac{2}{\pi}}e^{-ie(x-x')^{\mu}\int_0^1d\lambda A_{\mu}(\phi'+\lambda(\phi-\phi'))}\\
&\times\gamma^{\rho}\left[i\partial_{\rho}+e\int_0^1d\lambda\,\lambda F_{\rho\sigma}(\phi'+\lambda(\phi-\phi'))(x-x')^{\sigma}+m\right]\\
&\times \frac{e^{-i\tilde{m}(x,x')\sqrt{(x-x')^2}}}{\sqrt{\tilde{m}(x,x')\sqrt{(x-x')^2}}}\left[\frac{\tilde{m}(x,x')}{\sqrt{(x-x')^2}}+\frac{e}{2}\sigma^{\alpha\beta}\int_0^1d\lambda\,F_{\alpha\beta}(\phi'+\lambda(\phi-\phi'))\right].
\end{split}
\end{equation}
In the latter sub-case we recall that \cite{Gradshteyn_b_2000}
\begin{align}
\text{K}_{\nu}(z)\sim\sqrt{\frac{2}{\pi z}}e^{-z}  && z\to \infty,
\end{align}
such that, in the corresponding limit $(x-x')^2\to -\infty$, we have
\begin{equation}
\begin{split}
G(x,x')\sim &-\frac{i}{4\pi^2}\sqrt{\frac{\pi}{2}}e^{-ie(x-x')^{\mu}\int_0^1d\lambda A_{\mu}(\phi'+\lambda(\phi-\phi'))}\\
&\times\gamma^{\rho}\left[i\partial_{\rho}+e\int_0^1d\lambda\,\lambda F_{\rho\sigma}(\phi'+\lambda(\phi-\phi'))(x-x')^{\sigma}+m\right]\\
&\times \frac{e^{-\tilde{m}(x,x')\sqrt{-(x-x')^2}}}{\sqrt{\tilde{m}(x,x')\sqrt{-(x-x')^2}}}\left[\frac{\tilde{m}(x,x')}{\sqrt{-(x-x')^2}}-i\frac{e}{4}\sigma^{\alpha\beta}\int_0^1d\lambda\,F_{\alpha\beta}(\phi'+\lambda(\phi-\phi'))\right].
\end{split}
\end{equation}
As expected from causality considerations, the Green's function is highly oscillating for large time-like intervals and exponentially suppressed for large space-time intervals. In either case the typical ``length'' of the oscillation/exponential suppression is given by the effective, space-time dependent Compton length $1/\tilde{m}(x,x')\le\lambda_C$ (see also \cite{Hebenstreit_2011}).

\section{The Green's function of the Dirac equation in a plane wave and the Volkov propagator}

The Volkov states are the exact, analytical solutions of the Dirac equation in a plane wave \cite{Volkov_1935,Landau_b_4_1982}. The positive- and negative-energy Volkov states $U_s(\bm{p},x)$ and $V_s(\bm{p},x)$ can be classified by means of the asymptotic momentum quantum numbers $\bm{p}$ (and then the energy $\varepsilon =\sqrt{m^2+\bm{p}^2}$) and of the asymptotic spin quantum number $s=1,2$ in the remote past, i.e. for $t\to-\infty$. Following the general notation in \cite{Landau_b_4_1982}, these states are given by 
\begin{align}
\label{U_c}
U_s(\bm{p},x)&=\bigg[1+\frac{e\hat{n}\hat{A}(\phi)}{2p_-}\bigg]\text{e}^{i\left\{-(px)-\int_{-\infty}^{\phi}d\varphi\left[\frac{e(pA(\varphi))}{p_-}-\frac{e^2A^2(\varphi)}{2p_-}\right]\right\}}u_s(\bm{p}),\\
\label{V_c}
V_s(\bm{p},x)&=\bigg[1-\frac{e\hat{n}\hat{A}(\phi)}{2p_-}\bigg]\text{e}^{i\left\{(px)-\int_{-\infty}^{\phi}d\varphi\left[\frac{e(pA(\varphi))}{p_-}+\frac{e^2A^2(\varphi)}{2p_-}\right]\right\}}v_s(\bm{p}),
\end{align}
Here, we have introduced the on-shell four-momentum $p^{\mu}=(\varepsilon ,\bm{p})$ and the free, positive- and negative-energy spinors $u_s(\bm{p})$ and $v_s(\bm{p})$, respectively, normalized as $u^{\dag}_s(\bm{p})u_{s'}(\bm{p})=v^{\dag}_s(\bm{p})v_{s'}(\bm{p})=2\varepsilon \delta_{ss'}$ and such that $u^{\dag}_s(\bm{p})v_{s'}(-\bm{p})=0$. Even though the orthonormality of the Volkov states will be derived in the next section, since it is already known, we report here for the sake of convenience the final result (see, e.g., \cite{Ritus_1985}):
\begin{align}
\label{UU_VV_c}
\int d^3x\, U^{\dag}_s(\bm{p},x)U_{s'}(\bm{p}',x)=&\int d^3x\,V^{\dag}_s(\bm{p},x)V_{s'}(\bm{p}',x)=2\varepsilon (2\pi)^3\delta^{(3)}(\bm{p}-\bm{p}')\delta_{ss'},\\
\label{UV_VU_c}
\int d^3x\,U^{\dag}_s(\bm{p},x)V_{s'}(-\bm{p}',x)=&\int d^3x\,V^{\dag}_s(\bm{p},x)U_{s'}(-\bm{p}',x)=0.
\end{align}
Anticipating also that the positive- and negative-energy Volkov states are a complete set of states in the Hilbert space at hand, the Dirac field operator $\Psi(x)$ can be expanded as
\begin{equation}
\label{Psi}
\Psi(x)=\sum_s\int\frac{d^3p}{(2\pi)^3}\frac{1}{\sqrt{2\varepsilon }}\left[c_s(\bm{p})U_s(\bm{p},x)+d^{\dag}_s(\bm{p})V_s(\bm{p},x)\right],
\end{equation}
where $c_s(\bm{p})$ and $c^{\dag}_s(\bm{p})$ [$d_s(\bm{p})$ and $d^{\dag}_s(\bm{p})$] are the electron (positron) annihilation and creation operators, respectively, satisfying the anti-commutation relations $\{c_s(\bm{p}),c^{\dag}_{s'}(\bm{p}')\}=\{d_s(\bm{p}),d^{\dag}_{s'}(\bm{p}')\}=(2\pi)^3\delta^{(3)}(\bm{p}-\bm{p}')\delta_{ss'}$, with all other anti-commutators vanishing \cite{Peskin_b_1995}. 

Now, we recall that the Volkov propagator $G(x,x')$ is defined as (we already use the same symbol, which denotes the Green's function because below we will show that these two functions indeed coincide):
\begin{equation}
\label{VP_0}
G(x,x')=-i\langle 0|\mathcal{T}[\Psi(x)\bar{\Psi}(x')]|0\rangle,
\end{equation}
where $|0\rangle$ is the vacuum state corresponding to the electron/positron creation and annihilation operators introduced above, $\mathcal{T}$ is the time-ordering operator and $\bar{\Psi}(x)=\Psi^{\dag}(x)\gamma^0$ (the ``bar'' operation is analogously defined for any spinor). By using Eq. (\ref{Psi}) and the anti-commutation relations among the creation and annihilation operators, we obtain
\begin{equation}
\begin{split}
G(x,x')=&-i\theta(x^0-x^{\prime\,0})\sum_s\int\frac{d^3p}{(2\pi)^3}\frac{1}{2\varepsilon }U_s(\bm{p},x)\bar{U}_s(\bm{p},x')\\
&+i\theta(x^{\prime\,0}-x^0)\sum_s\int\frac{d^3p}{(2\pi)^3}\frac{1}{2\varepsilon }V_s(\bm{p},x)\bar{V}_s(\bm{p},x').
\end{split}
\end{equation}
It is convenient at this point to introduce the Ritus matrices $E(x,q)$ \cite{Ritus_1985} for an arbitrary (non-necessarily on-shell) four-momentum $q^{\mu}$ as
\begin{equation}
\label{E_xq}
E(x,q)=\bigg[1+\frac{e\hat{n}\hat{A}(\phi)}{2q_-}\bigg]\text{e}^{i\left\{-(qx)-\int_{-\infty}^{\phi}d\varphi\left[\frac{e(qA(\varphi))}{q_-}-\frac{e^2A^2(\varphi)}{2q_-}\right]\right\}},
\end{equation}
such that $U_s(\bm{p},x)=E(x,p)u_s(\bm{p})$ and $V_s(\bm{p},x)=E(x,-p)v_s(\bm{p})$ and \cite{Ritus_1985}
\begin{align}
\label{E_Ebar}
\int d^4x\, E(x,q)\bar{E}(x,q')=\int d^4x\, \bar{E}(x,q) E(x,q')=(2\pi)^4\delta^{(4)}(q-q'),\\
\label{Ebar_E}
\int \frac{d^4q}{(2\pi)^4} E(x,q)\bar{E}(x',q)=\int \frac{d^4q}{(2\pi)^4} \bar{E}(x,q)E(x',q)=\delta^{(4)}(x-x'),
\end{align}
with $\bar{E}(x,q)=\gamma^0E^{\dag}(x,q)\gamma^0$ (the ``bar'' operation on an arbitrary matrix or matrix operator is defined analogously). Since the free spinors $u_s(\bm{p})$ and $v_s(\bm{p})$ satisfy the relations $\sum_s u_s(\bm{p})\bar{u}_s(\bm{p})=\hat{p}+m$ and $\sum_s v_s(\bm{p})\bar{v}_s(\bm{p})=\hat{p}-m$ \cite{Landau_b_4_1982}, it is clear that
\begin{equation}
\label{VP_d3p}
\begin{split}
G(x,x')=&-i\theta(x^0-x^{\prime\,0})\int\frac{d^3p}{(2\pi)^3}\frac{1}{2\varepsilon }E(x,p)(\hat{p}+m)\bar{E}(x',p)\\
&+i\theta(x^{\prime\,0}-x^0)\int\frac{d^3p}{(2\pi)^3}\frac{1}{2\varepsilon }E(x,-p)(\hat{p}-m)\bar{E}(x',-p).
\end{split}
\end{equation}
In order to establish the identity between the Green's function of the Dirac operator in a plane wave and the Volkov propagator, we first show that the latter can be written as
\begin{equation}
\label{VP}
G(x,x')=\int\frac{d^4q}{(2\pi)^4}E(x,q)\frac{\hat{q}+m}{q^2-m^2+i0}\bar{E}(x',q).
\end{equation}
The identity is indeed easily proved in an analogous way as in the free case (see, e.g., \cite{Landau_b_4_1982}) once one observes that
\begin{enumerate}
\item the divergence (pole) in $q^0$ at $q_-=q^0-q_n=0$ can be integrated, as we have explicitly shown in the previous section, due to the corresponding fast oscillations of the exponential function whose argument also diverges at $q_-=0$ (the integration over $q_-$ there is reduced to an integration over the proper time $s$);
\item when the integral in $q^0$ is evaluated on a large semicircle either with $\text{Im}(q^0)>0$ or with $\text{Im}(q^0)<0$, the additional terms in the phase related to the plane wave with respect to the free case tend to zero as they are inversely proportional to $q_-=q^0-q_n$.
\end{enumerate}
Accounting now for these remarks, one can perform the integral in $q^0$ in Eq. (\ref{VP}) exactly as in the vacuum case and the expression in Eq. (\ref{VP_d3p}) is obtained. Finally, the equivalence between the Green's function of the previous section and the Volkov propagator is established once we notice that (see, e.g., \cite{Ritus_1985,Meuren_2013})
\begin{equation}
\label{PEEp}
\gamma^{\mu}[i\partial_{\mu}-eA_{\mu}(\phi)]E(x,q)=E(x,q)\hat{q},
\end{equation}
where it is understood that no other functions of $x^{\mu}$ are present on the right of $E(x,q)$ in the left hand side. It is important to stress that in order to prove that the function $G(x,x')$ in Eq. (\ref{VP}) satisfies the Green's function equation like Eq. (\ref{Eq_G_b}) with $A_b^{\mu}(x)=A^{\mu}(\phi)$, only the properties in Eqs. (\ref{Ebar_E}) and (\ref{PEEp}) are necessary. These properties can be easily proved directly from the expression of the Ritus matrices without relying, in particular, on the completeness of the Volkov states mentioned below Eq. (\ref{VP_0}).

As a side remark, we notice that all the above results on the Ritus matrices can be easily obtained within the operator technique by introducing the operator
\begin{equation}
\label{E}
E=\bigg[1-\frac{e\hat{n}\hat{A}(\Phi)}{2P_T}\bigg]\text{e}^{i\int_{-\infty}^{\Phi}d\varphi\left[\frac{e(PA(\varphi))}{P_T}-\frac{e^2A^2(\varphi)}{2P_T}\right]}
\end{equation}
such that $E(x,q)=\langle x|E|q\rangle$ [recall that $P_T|q\rangle=-q_-|q\rangle$, that $\langle x|q\rangle=\exp(-i(qx))$ and that in the Lorenz gauge $(PA(\Phi))=(A(\Phi)P)$]. The relations (\ref{E_Ebar}) and (\ref{Ebar_E}) among the Ritus matrices correspond to the identities $E\bar{E}=\bar{E}E=I$, with $I$ being the unity operator, whereas the corresponding operator equation of Eq. (\ref{PEEp}) reads $\hat{\Pi}E=E\hat{P}$, which can all be proved directly starting from Eq. (\ref{E}). Finally, the Volkov propagator in Eq. (\ref{VP}) and its equivalence with the Green's function simply corresponds to the operator equation $G=E(\hat{P}-m+i0)^{-1}\bar{E}=(\hat{\Pi}-m+i0)^{-1}$, with the second identity following from the identity $\hat{\Pi}E=E\hat{P}$.

\section{A proof of the completeness and of the orthonormality of the Volkov states}

As we have mentioned in the Introduction, we would like to present alternative and relatively straightforward proofs of the completeness and the orthonormality of the Volkov states at a fixed time. We have also recalled that these properties have been already proved in \cite{Boca_2011} and in \cite{Ritus_1985,Zakowicz_2005}, respectively. In our opinion, however, the proofs below shed light on some interesting features, which are worth being noticed.

The completeness of the Volkov states at a fixed time $t$ is expressed by the identity \cite{Boca_2011}
\begin{equation}
\label{Compl_V}
\sum_s\int\frac{d^3p}{(2\pi)^3}\frac{1}{2\varepsilon }\left[U_s(\bm{p},t,\bm{x})U^{\dag}_s(\bm{p},t,\bm{x}')+V_s(\bm{p},t,\bm{x})V^{\dag}_s(\bm{p},t,\bm{x}')\right]=\delta^{(3)}(\bm{x}-\bm{x}').
\end{equation}
Now, we start from the expression of the Green's function $G(x,x')$ of the Dirac equation in a plane wave in Eq. (\ref{VP}), which, by exploiting the two remarks below that equation, can be written in the form as in Eq. (\ref{VP_d3p}). Also, by using the properties of the Ritus matrices in Eq. (\ref{Ebar_E}) and in Eq. (\ref{PEEp}), we have shown that $\{\gamma^{\mu}[i\partial_{\mu}-eA_{\mu}(\phi)]-m\}G(x,x')=\delta^{(4)}(x-x')$, i.e., that
\begin{equation}
\begin{split}
\delta^{(4)}(x-x')=&\{\gamma^{\mu}[i\partial_{\mu}-eA_{\mu}(\phi)]-m\}\left[-i\theta(x^0-x^{\prime\,0})\int\frac{d^3p}{(2\pi)^3}\frac{1}{2\varepsilon }E(x,p)(\hat{p}+m)\bar{E}(x',p)\right.\\
&+i\theta(x^{\prime\,0}-x^0)\int\frac{d^3p}{(2\pi)^3}\frac{1}{2\varepsilon }E(x,-p)(\hat{p}-m)\bar{E}(x',-p)\bigg]=\\
&=\{\gamma^{\mu}[i\partial_{\mu}-eA_{\mu}(\phi)]-m\}\bigg[-i\theta(x^0-x^{\prime\,0})\sum_s\int\frac{d^3p}{(2\pi)^3}\frac{1}{2\varepsilon }U_s(\bm{p},x)\bar{U}_s(\bm{p},x')\\
&+i\theta(x^{\prime\,0}-x^0)\sum_s\int\frac{d^3p}{(2\pi)^3}\frac{1}{2\varepsilon }V_s(\bm{p},x)\bar{V}_s(\bm{p},x')\bigg],
\end{split}
\end{equation}
where in the second equality we have used the expression of the Volkov states in terms of the Ritus matrices [see the relations between Eq. (\ref{E_xq}) and Eq. (\ref{E_Ebar})]. At this point, the derivative $\partial_{\mu}$ acts both on the theta-functions and on the Volkov states. Since the Volkov states are solutions of the Dirac equation in a plane wave, we obtain
\begin{equation}
\delta^{(4)}(x-x')=\delta(x^0-x^{\prime\,0})\gamma^0\sum_s\int\frac{d^3p}{(2\pi)^3}\frac{1}{2\varepsilon }[U_s(\bm{p},x)\bar{U}_s(\bm{p},x')+V_s(\bm{p},x)\bar{V}_s(\bm{p},x')],
\end{equation}
which, after multiplying it by $\gamma^0$ once from the left and once from the right and integrating with respect to either $x^0$ or to $x^{\prime\,0}$, implies the completeness relation in Eq. (\ref{Compl_V}) with $x^0=x^{\prime\,0}=t$. It is worth repeating [see also the discussion below Eq. (\ref{PEEp})], that the equivalence between Eq. (\ref{VP_d3p}) and Eq. (\ref{VP}) as well as the properties of the Ritus matrices in Eq. (\ref{Ebar_E}) and in Eq. (\ref{PEEp}), can be proved independently of the completeness of the Volkov states, which was used in Eq. (\ref{Psi}) only to show the equivalence between the Green's function of the Dirac equation in a plane wave and the quantity $-i\langle 0|\mathcal{T}[\Psi(x)\bar{\Psi}(x')]|0\rangle$, i.e., the dressed Feynman propagator of the Dirac field.

We pass now to prove the orthonormality of the Volkov states at a given time in Eqs. (\ref{UU_VV_c}) and (\ref{UV_VU_c}). In fact, in order to follow the method outlined below, it is more convenient to start by assuming a finite quantization volume, to be a large cube of side $L$ and volume $V=L^3$ centered at the origin of the coordinates. In this way, by imposing suitable periodic boundary conditions (see below), the momenta quantum numbers become discrete and the Volkov states can be normalized to a finite value. Only at the end of any actual calculation the limit $L\to\infty$ has to be be taken.

Since the Volkov states in the case of a finite quantization volume differ from those in Eqs. (\ref{U_c}) and (\ref{V_c}) by a factor $1/\sqrt{V}$, they will be indicated by different symbols as
\begin{align}
\label{U}
U_{\bm{p},s}(x)&=\bigg[1+\frac{e\hat{n}\hat{A}(\phi)}{2p_-}\bigg]\text{e}^{i\left\{-(px)-\int_{-\infty}^{\phi}d\varphi\left[\frac{e(pA(\varphi))}{p_-}-\frac{e^2A^2(\varphi)}{2p_-}\right]\right\}}\frac{u_s(\bm{p})}{\sqrt{V}},\\
\label{V}
V_{\bm{p},s}(x)&=\bigg[1-\frac{e\hat{n}\hat{A}(\phi)}{2p_-}\bigg]\text{e}^{i\left\{(px)-\int_{-\infty}^{\phi}d\varphi\left[\frac{e(pA(\varphi))}{p_-}+\frac{e^2A^2(\varphi)}{2p_-}\right]\right\}}\frac{v_s(\bm{p})}{\sqrt{V}}.
\end{align}
The orthonormality at a fixed time $t$ corresponding to Eqs. (\ref{UU_VV_c})-(\ref{UV_VU_c}) reads
\begin{align}
\label{UU_VV}
\int_V d^3x\, U^{\dag}_{\bm{p},s}(x)U_{\bm{p}',s'}(x)=&\int_V d^3x\,V^{\dag}_{\bm{p},s}(x)V_{\bm{p}',s'}(x)=2\varepsilon \delta_{\bm{p}\bm{p}'}\delta_{ss'},\\
\label{UV_VU}
\int_V d^3x\,U^{\dag}_{\bm{p},s}(x)V_{-\bm{p}',s'}(x)=&\int_V d^3x\,V^{\dag}_{\bm{p},s}(x)U_{-\bm{p}',s'}(x)=0.
\end{align}
Now, for the sake of definiteness, we assume that the plane wave propagates along the positive $z$ direction. Thus, it is clear that the periodic boundary conditions along the $x$ direction [$U_{\bm{p},s}(x^0,-L/2,x^2,x^3)=U_{\bm{p},s}(x^0,L/2,x^2,x^3)$ and $V_{\bm{p},s}(x^0,-L/2,x^2,x^3)=V_{\bm{p},s}(x^0,L/2,x^2,x^3)$] and the $y$ direction [$U_{\bm{p},s}(x^0,x^1,-L/2,x^3)=U_{\bm{p},s}(x^0,x^1,L/2,x^3)$ and $V_{\bm{p},s}(x^0,x^1,-L/2,x^3)=V_{\bm{p},s}(x^0,x^1,L/2,x^3)$] imply the usual conditions $p_xL=2\ell_x\pi$ and $p_yL=2\ell_y\pi$, with $\ell_x,\ell_y=0,\pm 1,\pm 2,...$ as in the vacuum case. Concerning the $z$ direction, we observe that since the limit $L\to \infty$ is ultimately taken, we can assume that the external field is defined as a given, arbitrary function between, for example, $\phi=-L/2$ and $\phi=L/2$ and it is extended in a periodic way beyond those limits, i.e., $A^{\mu}(t-L/2)=A^{\mu}(t+L/2)$ for any $t$. Thus, the remaining periodic boundary condition for positive- and negative-energy states implies the conditions [see Eqs. (\ref{U}) and (\ref{V})]
\begin{align}
\label{PBC_U_z}
p_zL+\int_{t-L/2}^{t+L/2}d\phi\left[\frac{e(pA)}{p_-}-\frac{e^2A^2}{2p_-}\right]= p_zL+\int_{-L/2}^{L/2}d\phi\left[\frac{e(pA)}{p_-}-\frac{e^2A^2}{2p_-}\right]=2\ell_z\pi,\\
\label{PBC_V_z}
p_zL-\int_{t-L/2}^{t+L/2}d\phi\left[\frac{e(pA)}{p_-}-\frac{e^2A^2}{2p_-}\right]= p_zL-\int_{-L/2}^{L/2}d\phi\left[\frac{e(pA)}{p_-}-\frac{e^2A^2}{2p_-}\right]=2\ell_z\pi,
\end{align}
respectively, with $\ell_z=0,\pm 1,\pm 2,...$. By considering realistic plane-wave fields such that the integrals $\int_{-\infty}^{\infty}d\phi\,A^{\mu}(\phi)$ and $\int_{-\infty}^{\infty}d\phi\,A^2(\phi)$ are finite and having in mind the limit $L\to \infty$, we can also conclude that the density of Volkov states still corresponds to the formal substitution rule $\sum_{\bm{\ell}}\to V\int d^3\bm{p}/(2\pi)^3$ as in the vacuum case. This discussion also clarifies why for a monochromatic plane wave with amplitude $A_0^{\mu}$ and angular frequency $\omega_0$, i.e., in the case of linear polarization, $A^{\mu}(\phi)=A_0^{\mu}\cos(\omega_0\phi)$, periodic boundary conditions on the electron ``four-quasimomentum'' $p^{\mu}-e^2A_0^2n^{\mu}/4p_-$ have to be enforced both for positive- and negative-energy Volkov states \cite{Ritus_1985}. At this point, by indicating as $\Psi_{\bm{p},s}(x)$ any either positive- or negative-energy Volkov state, we can first easily show that
\begin{equation}
\label{Cont}
\partial_{\mu}[\bar{\Psi}_{\bm{p},s}(x)\gamma^{\mu}\Psi_{\bm{p}',s'}(x)]=0
\end{equation}
as a consequence of the fact that the states $\Psi_{\bm{p},s}(x)$ and $\Psi_{\bm{p}',s'}(x)$ are solutions of the Dirac equation in the plane wave, i.e.,
\begin{align}
\label{DE}
\{\gamma^{\mu}[i\partial_{\mu}-eA_{\mu}(\phi)]-m\}\Psi_{\bm{p}',s'}(x)&=0,\\
\label{bar_DE}
\bar{\Psi}_{\bm{p},s}(x)\{\gamma^{\mu}[-i\overleftarrow{\partial}_{\mu}-eA_{\mu}(\phi)]-m\}&=0.
\end{align}
Eq. (\ref{Cont}), in fact, holds as it is proportional to the difference between Eq. (\ref{DE}) multiplied on the left by $\bar{\Psi}_{\bm{p},s}(x)$ and Eq. (\ref{bar_DE}) multiplied on the right by $\Psi_{\bm{p}',s'}(x)$. Now, we fix a four-dimensional volume $\Omega$ consisting of the three-dimensional volume $V$ extended along the time direction from $x^0\to -\infty$ to $x^0=t$, with $t$ being an arbitrary time, where the presence of the plane wave cannot be neglected. It is clear that $\int_{\Omega}d^4x\,\partial_{\mu}[\bar{\Psi}_{\bm{p},s}(x)\gamma^{\mu}\Psi_{\bm{p}',s'}(x)]=0$ and then, by applying the Gauss theorem in four dimensions \cite{Landau_b_2_1975}, that
\begin{equation}
\label{Orth}
\int_{\Sigma}d\Sigma_{\mu}\bar{\Psi}_{\bm{p},s}(x)\gamma^{\mu}\Psi_{\bm{p}',s'}(x)=0,
\end{equation}
where $\Sigma$ is the three-dimensional hyper-surface enclosing the four-dimensional volume $\Omega$. Given the shape of the four-dimensional volume $\Omega$, Eq. (\ref{Orth}) can be written as
\begin{equation}
\label{Orth_2}
\begin{split}
0=&\int_Vd^3x\,\bar{\Psi}_{\bm{p},s}(t,\bm{x})\gamma^0\Psi_{\bm{p}',s'}(t,\bm{x})-\int_Vd^3x\,\lim_{x^0\to-\infty}\bar{\Psi}_{\bm{p},s}(x)\gamma^0\Psi_{\bm{p}',s'}(x)\\
&-\int_{\Sigma_1}dx^0dx^2dx^3\,\bar{\Psi}_{\bm{p},s}(x_{1,+})\gamma^1\Psi_{\bm{p}',s'}(x_{1,+})+\int_{\Sigma_1}dx^0dx^2dx^3\,\bar{\Psi}_{\bm{p},s}(x_{1,-})\gamma^1\Psi_{\bm{p}',s'}(x_{1,-})\\
&-\int_{\Sigma_2}dx^0dx^1dx^3\,\bar{\Psi}_{\bm{p},s}(x_{2,+})\gamma^2\Psi_{\bm{p}',s'}(x_{2,+})+\int_{\Sigma_2}dx^0dx^1dx^3\,\bar{\Psi}_{\bm{p},s}(x_{2,-})\gamma^2\Psi_{\bm{p}',s'}(x_{2,-})\\
&-\int_{\Sigma_3}dx^0dx^1dx^2\,\bar{\Psi}_{\bm{p},s}(x_{3,+})\gamma^3\Psi_{\bm{p}',s'}(x_{3,+})+\int_{\Sigma_3}dx^0dx^1dx^2\,\bar{\Psi}_{\bm{p},s}(x_{3,-})\gamma^3\Psi_{\bm{p}',s'}(x_{3,-}),
\end{split}
\end{equation}
where the hyper-surface $\Sigma_j$, with $j=1,2,3$, is obtained by extending from $x^0\to -\infty$ to $x^0=t$ a square of side $L$ centered on the origin of the plane perpendicular to the $j$th direction, and where $x_{1,\pm}=(x^0,\pm L/2,x^2,x^3)$, $x_{2,\pm}=(x^0,x^1,\pm L/2,x^3)$, and $x_{3,\pm}=(x^0,x^1,x^2,\pm L/2)$. Now, it is clear that the first line of Eq. (\ref{Orth_2}) gives already the desired result. In fact, the first term is exactly the scalar product of two Volkov states at a given time $x^0=t$, where the presence of the plane wave cannot be neglected. The second term, instead, coincides with the same scalar product but for $x^0\to-\infty$, where the Volkov states coincide with the free states, which indeed fulfill the wanted orthonormality conditions [see Eqs. (\ref{UU_VV})-(\ref{UV_VU})]. Analogous conclusions can be drawn for the remaining cases also involving Volkov states with negative energy. In this way, the Volkov states fulfill Eqs. (\ref{UU_VV})-(\ref{UV_VU}), once we proof that the remaining contributions in Eq. (\ref{Orth_2}) vanish. Since the plane wave propagates along the positive $z$ direction, it is clear that the two contributions on the second line as well as the two on the third line cancel each other because of the periodic boundary conditions. The same is true also for the last line and, for the sake of definiteness, we will explicitly prove it only for two positive-energy Volkov states. By recalling the periodicity conditions on the plane-wave four-vector potential, the last line of Eq. (\ref{Orth_2}) becomes
\begin{equation}
\begin{split}
&\int_{\Sigma_3}dx^0dx^1dx^2\bar{U}_{\bm{p},s}(x_{3,-})\gamma^3U_{\bm{p}',s'}(x_{3,-})-\int_{\Sigma_3}dx^0dx^1dx^2\bar{U}_{\bm{p},s}(x_{3,+})\gamma^3U_{\bm{p}',s'}(x_{3,+})\\
&=\delta_{\bm{p}_{\perp}\bm{p}'_{\perp}}\int_{-\infty}^t \frac{dx^0}{L}\bar{u}_s(\bm{p})\left[1-\frac{e\hat{n}\hat{A}(x^0+L/2)}{2p_-}\right]\gamma^3\left[1+\frac{e\hat{n}\hat{A}(x^0+L/2)}{2p'_-}\right]u_{s'}(\bm{p}')\\
&\qquad\times\text{e}^{i\left\{(\varepsilon -\varepsilon')x^0+(p_z-p'_z)\frac{L}{2}-\int_{-\infty}^{x^0+L/2}d\varphi\left[\frac{e(p'A(\varphi))}{p'_-}-\frac{e(pA(\varphi))}{p_-}-\frac{e^2A^2(\varphi)}{2}\left(\frac{1}{p'_-}-\frac{1}{p_-}\right)\right]\right\}}\\
&\quad-\delta_{\bm{p}_{\perp}\bm{p}'_{\perp}}\int_{-\infty}^t \frac{dx^0}{L}\bar{u}_s(\bm{p})\left[1-\frac{e\hat{n}\hat{A}(x^0-L/2)}{2p_-}\right]\gamma^3\left[1+\frac{e\hat{n}\hat{A}(x^0-L/2)}{2p'_-}\right]u_{s'}(\bm{p}')\\
&\qquad\times\text{e}^{i\left\{(\varepsilon -\varepsilon')x^0-(p_z-p'_z)\frac{L}{2}-\int_{-\infty}^{x^0-L/2}d\varphi\left[\frac{e(p'A(\varphi))}{p'_-}-\frac{e(pA(\varphi))}{p_-}-\frac{e^2A^2(\varphi)}{2}\left(\frac{1}{p'_-}-\frac{1}{p_-}\right)\right]\right\}}\\
&=\delta_{\bm{p}_{\perp}\bm{p}'_{\perp}}\int_{-\infty}^t \frac{dx^0}{L}\bar{u}_s(\bm{p})\left[1-\frac{e\hat{n}\hat{A}(x^0+L/2)}{2p_-}\right]\gamma^3\left[1+\frac{e\hat{n}\hat{A}(x^0+L/2)}{2p'_-}\right]u_{s'}(\bm{p}')\\
&\qquad\times\text{e}^{i\left\{(\varepsilon -\varepsilon')x^0+(p_z-p'_z)\frac{L}{2}-\int_{-\infty}^{x^0+L/2}d\varphi\left[\frac{e(p'A(\varphi))}{p'_-}-\frac{e(pA(\varphi))}{p_-}-\frac{e^2A^2(\varphi)}{2}\left(\frac{1}{p'_-}-\frac{1}{p_-}\right)\right]\right\}}\\
&\qquad\times\left\langle 1-\text{e}^{i\left\{-(p_z-p'_z)L+\int_{x^0-L/2}^{x^0+L/2}d\varphi\left[\frac{e(p'A(\varphi))}{p'_-}-\frac{e(pA(\varphi))}{p_-}-\frac{e^2A^2(\varphi)}{2}\left(\frac{1}{p'_-}-\frac{1}{p_-}\right)\right]\right\}}\right\rangle,
\end{split}
\end{equation}
which vanishes due to Eq. (\ref{PBC_U_z}). Since the remaining cases involving the negative-energy Volkov states can be addressed analogously, we can conclude that the orthonormality properties in Eqs. (\ref{UU_VV})-(\ref{UV_VU}) hold. Finally, by performing the limit $L\to\infty$, one obtains the orthonormality properties in Eqs. (\ref{UU_VV_c})-(\ref{UV_VU_c}).

\section{Conclusions}

In this paper we have provided alternative and relatively simple proofs of the completeness and the orthonormality of the Volkov states at a fixed time. Whereas the orthonormality relations have been proved by relying on a geometrical argument based on the Gauss theorem in four dimensions, the proof of the completeness of the Volkov states exploited some properties of the Green's function $G(x,x')$ of the Dirac operator in a plane wave and the Volkov propagator. Moreover, an explicit expression of $G(x,x')$ has been reported in terms of special functions (modified Bessel functions and Hankel functions) and their derivatives, and its equivalence with the Volkov propagator has been proved explicitly. Also, some asymptotic expressions for small and large (both spacelike and timelike) four-dimensional square distances $(x-x')^2$ have been derived. Special attention has been also devoted to the transformation properties of the Green's function $G(x,x')$ under general gauge transformation (which do not necessarily keep the four-vector potential be dependent only on a single phase variable) and the related invariance of the dressed mass.

\acknowledgments{The author would like to thank A. Angioi, S. Bragin, R. Zh. Shaisultanov, and O. D. Skoromnik for insightful discussions and for reading the manuscript.}

\end{document}